\documentclass[12pt]{article}
\usepackage{amssymb}
\usepackage[dvips]{graphicx}
\usepackage{graphics}
\usepackage{epsfig}
\usepackage{latexsym}
\usepackage[figuresright]{rotating}
\begin{document}
\begin{center}
{\Large \textbf{Low-energy  electromagnetic characteristics of the
$\pi$-meson in the covariant three-dimensional approach}}
\\
\textbf{\large N.V.Maksimenko} \footnote{\textsf{E-mail}:\textbf{
MAKSIMENKO@GSU.UNIBEL.BY}}\\
\textbf{\large O.M.Deryuzhkova} \footnote{\textsf{E-mail}:\textbf{
DOM@GSU.UNIBEL.BY}}\\
Gomel State University, Physics Department, \\
Gomel, Belarus
\end{center}
\large
\begin{abstract}
The inverse time ordered  Green's function is defined in a covariant
three-dimensional formalism using the diagram approach.
The equation of motion of a two-quark composite system in an external
electromagnetic field up to the second order in the field
strength is obtained.
The
polarizability of the $ \pi$-meson
is calculated using this formalism. The mean square radius of the
two-quark system is calculated from the wave functions of the
system.
 In addition, radiative corrections are calculated for
both the  polarizability and the mean square radius.
\end{abstract}

Large attention is recently devoted to the electromagnetic
characteristics of hadrons described as composite systems \cite{usa1}.
The polarizability is an particular important observable among
the set of the electromagnetic characteristics, since it is very
sensitive to the model used to describe the interactions of quarks
\cite{usa2, usa6}.
Numerical evaluations using these models have shown that for
example the  polarizability of $\pi$-mesons probes the
contributions of relativistic corrections.

The description of the electromagnetic behaviour of relativistic
coupled systems within the framework of relativistic quantum
mechanics faces large difficulties stipulated by the necessity of
the correct treatment of the motion of the center of mass and the
relative motion of the system due to the interaction with an
electromagnetic field \cite{usa7}.
For the description of coupled systems on the basis of the diagram
approach the covariant four-dimensional Bethe-Salpeter equations
are obtained. However, their practical usage is accompanied due to
a lot of difficulties \cite{usa5}. Therefore for the solution
of a particular problem usually the three dimensional
Bethe-Salpeter equations are used. In these equations vertex state
functions
 of the system of two
particles are projected onto the surface of equal time in the
system of center of mass \cite{usa5,usa3}. Such an approach has
been successfully performed for the description of bound states of
particles in external electromagnetic fields \cite{usa2}, \cite{usa5}, \cite{usa9}.

In quantum field theory the formula for the electrical
polarizability of $\pi$-mesons, $\alpha$, is obtained using
low-energy theorems of the Compton scattering amplitude, and has
the following form \cite{usa8}:
\begin{equation}
\alpha = \alpha_0 + \Delta \alpha, \label{usa1}
\end{equation}
where
\[ \alpha _0=2\sum _{s\neq n}\frac{\left| \left\langle \frac
{d}{dp}\right. \left\langle n\stackrel{\rightarrow} {p}\right|
\hat{J}_0\left| s0\right\rangle \right|^2
_{\stackrel{\rightarrow} {p}=0}} {E_{s0}-M}
\]
and
\[
\Delta \alpha =\frac{e^2\left\langle r^2\right\rangle }{3M}.
\]
In expression (\ref{usa1}) $\hat{J}_0$ is the zero component of
the operator of a 4-current vector, $M$ is the mass of a ground
state of the coupled system, $E_{s0}$ is the energy of the system
which is distinct from ground states, $\vec p$ is the momentum of
$n$-th exited state of the system and  $\left\langle
r^2\right\rangle$ is the mean square charge radius of the
$\pi$-meson. As can be seen from eqn. (\ref{usa1}), the summation
of all contributions from the corresponding matrix elements of the
currents for transitions to intermediate states is necessary.
Since this is practically impossible large uncertainties in the
estimation of $\alpha_0$ remain. It is well known that $\alpha _0$
can be calculated precisely for a nonrelativistic two-particle
coupled system, where the latter is described by model dependent
potentials, using the nonrelativistic equations of motion in an
external electromagnetic field \cite{usa8, usa4}.

In this paper, the equation of motion of a two-quark coupled
relativistic system in an external electromagnetic field is
obtained on the basis of the diagram approach with equal time
approximation of the vertex functions for meson-quark
interactions. This equation is solved using perturbation theory.
From the result the polarizability of the relativistic system is
obtained in the same way as for the electrical polarizability of
the hydrogen atom in nonrelativistic quantum mechanics \cite{usa8,
usa4}.
 Let's now calculate the mean
square charge radius of the $\pi$-meson in the composite quark
model. The quarks are assumed to be pointlike charged scalars and
the vertex function of the meson-quark interaction is not
dependent on the energy of the system.

In this case the form factor of the $\pi$-meson in the Breit frame
can be written as \cite{usa9}:
$$
F(\vec {k}^2)=\pi \int d\stackrel{\rightarrow} {p}_1 d
\stackrel{\rightarrow}{ p}_{1}^{\prime} d\stackrel{\rightarrow}{
p}_2 d\stackrel{\rightarrow }{ p}_{2}^{\prime} \Gamma _2 (
\stackrel{\rightarrow} {p}_{1}^{\prime}, \stackrel{\rightarrow}
{p}_{2}^{\prime} )$$
\begin{equation}
\cdot{\cal {G}}^{(1)} ( \stackrel{\rightarrow} {p}_{1}^{\prime},
\stackrel{\rightarrow} {p}_{2}^{\prime}| \stackrel{\rightarrow
}{p}_1, \stackrel{\rightarrow} {p}_2 ) \Gamma _1(
\stackrel{\rightarrow} {p}_1, \stackrel{\rightarrow} {p}_2 ),
\label{usa2}
\end{equation}
Where $\Gamma _1$ and $\Gamma _2$ are the meson-quark vertex
functions, ${\cal {G}}^{(1)}$ is the Green's function of the
interaction of the electromagnetic fields with the two-particle
system in the first order of the coupling constant,
$\stackrel{\rightarrow}{ p}_n$ and $ \stackrel{\rightarrow }{
p}_{n}^{\prime}$ are the momenta of particles before and after
interaction. Using the methods of the diagram approach it can be
shown that
$$ {\cal
{G}}^{(1)}={\cal {G}}_1^{(1)}+{\cal {G}}_2^{(1)},
$$
where the Green's function for the interaction between a photon and a
quark with a charge $e_1$ is given by:
\begin{equation}
{\cal {G}}_1^{(1)}=\frac {\pi} {i} \frac{\left( 2\pi \right)
^6}{2E_1E_2} \frac{\left( -4\right) e_1E_1a_{01}p_0 \left( \Sigma
+ \Sigma^{~\prime} \right)} {\left( \Sigma^2-p_0^2 \right)\left(
\Sigma^{~\prime 2} -p_0^2\right)\left( E_1+E_1^{\prime} \right)} ,
\label{usa4}
\end{equation}
and the Green's
function ${\cal {G}}_2^{(1)}$ for the interaction of a photon with a quark of a
charge $e_2$ by replacing
$e_1\rightarrow e_2$, $E_1\leftrightarrow E_2$,
$E_1^{\prime} \leftrightarrow E_2^{\prime}$.  In expression
(\ref{usa4}) $\Sigma=E_1+E_2$,
$\Sigma^{\prime}=E_1^{\prime}+E_2$ and $a_{01}$ is the
scalar potential of the electromagnetic field.

Let's now replace
in expression (\ref{usa2}) the vertex functions by
wave functions $\Psi$  of the two-quark system. For this purpose
we use the inverse Green's function of two noninteracting
particles:
\begin{equation}
{\cal {G}}^{(0)-1}=\frac {i}{\left( 2\pi \right) ^6}
\frac{E_1E_2\left( \Sigma^2- p_0^2\right)}{\pi\Sigma }.
 \label{usa5}
\end{equation}
The vertex function is expressed through the wave function as
follows:
\begin{equation}
\Gamma = {\cal {G}}^{(0)-1} \Psi.
 \label{usa6}
\end{equation}
The wave function in expression (\ref{usa6}) satisfies the
equation:
\begin{equation}
 \left( p_0^2 - \Sigma^2\right)\Psi \left
 (\stackrel{\rightarrow }{p}_1, \stackrel{\rightarrow} {p}_2 \right)=
 \int d\stackrel{\rightarrow}{ p}_{1}^{\prime} d\stackrel
 {\rightarrow }{ p}_{2}^{\prime} I \left(\stackrel{\rightarrow }{p}_1,
 \stackrel{\rightarrow} {p}_2;
\stackrel{\rightarrow} {p}_{1}^{\prime}, \stackrel{\rightarrow}
{p}_{2}^{\prime} \right) \Psi\left( \stackrel{\rightarrow}
{p}_{1}^{\prime}, \stackrel{\rightarrow} {p}_{2}^{\prime} \right),
 \label{usa7}
\end{equation}
Where $I$ is the kernel of the interaction. Substituting (\ref{usa5})
and (\ref{usa6}) into equation (\ref{usa2}) and
taking into account the decomposition
\begin{equation}
F(\vec {k}^2)\simeq 1 - \frac{1}{6}\left\langle r^2\right\rangle
\vec {k}^2,
\end{equation}
we get for $\left\langle
r^2\right\rangle$ the expression
\begin{equation}
\left\langle r^2\right\rangle =\int d\stackrel{\rightarrow
}{q}\Psi^{+}\left( \stackrel{\rightarrow }{q}\right) \left\{-
\frac 14 \stackrel{\rightarrow }{\partial }_q^2+\frac
3{64}\frac{\stackrel{ \rightarrow }{q}^2}{E_q^4}\right\}
\Psi\left( \stackrel{ \rightarrow }{q}\right), \label{usa9}
\end{equation}
where $\Psi $ is the wave function of the two-quark coupled
system, $
\vec q$ is the relative two-quark momentum and
$E_q^4=(\stackrel{\rightarrow}{q}^2+~m^2)^2$. In equation
(\ref{usa9}) the first term equals the nonrelativistic
expression obtained for $ \left\langle r^2\right\rangle $ and the
second represents the relativistic correction.

For the calculation of the polarizability of the two-quark system using
wave functions the  equation of motion of the system in
the electromagnetic field is necessary.
For this purpose we determine an
inverse Green's function in the second order on the value of
strength of the electromagnetic field. From the relation
\begin{equation}
{\cal {G}}^{-1} {\cal {G}}=1
\end{equation}
follows that the inverse Green's function ${\cal {G}}^{-1}$ can be
represented as a decomposition as follows:
\begin{equation}
{\cal {G}}^{-1} =b^{(0)}+ b^{(1)} + b^{(2)}. \label{usa12}
\end{equation}
In this expression
$$
b^{(0)}={\cal {G}}^{(0)-1},~~~ b^{(1)}= - {\cal {G}}^{(0)-1}{\cal
{G}}^{(1)}{\cal {G}}^{(0)-1},$$
\begin{equation}
b^{(2)}={\cal {G}}^{(0)-1}{\cal {G}}^{(1)}{\cal {G}}^{(0)-1}{\cal
{G}}^{(1)}{\cal {G}}^{(0)-1}- {\cal {G}}^{(0)-1}{\cal
{G}}^{(2)}{\cal {G}}^{(0)-1}=b_1^{(2)}+b_2^{(2)}. \label{usa13}
\end{equation}
The functions ${\cal {G}}^{(0)}, {\cal {G}}^{(1)}$ and ${\cal
{G}}^{(2)}$ are the Green's functions of the zero's, first and
second orders, respectively, of the electromagnetic field strength
$\vec {\cal {E}}$. These functions we calculate using the diagram
approach in the time ordered approximation. As follows from
expressions (\ref{usa13}), the sum contributions of the diagrams
describing the process, when one quark interacts with the photon,
and the another radiates, is equal to zero. Thus, the inverse
Green's function (\ref{usa12}), and, therefore, the equation of
motion are defined only by the contribution of amplitudes from the
diagrams of Compton scattering photons on separate quarks. This
results from the relativistic momentum approximation of the
interaction between photons with the composite system.

As follows from expression for a Green function (\ref{usa12}) the
equation of motion of the system consisting of two charged
particles in the external electrostatic field takes the form:
\begin{equation}
\left({\cal {G}}^{-1}-V \right)\Psi =0. \label{usa15}
\end{equation}
 The following denotations are used:
$$
{\cal {G}}^{-1}\Psi = \left\{ \left( E_1+E_2-p_0\right)
+ie_1\stackrel{\rightarrow }{{\cal{E}}} \stackrel{\rightarrow
}{\nabla }_{p_1}+ie_2\stackrel{\rightarrow }{
{\cal{E}}}\stackrel{\rightarrow }{\nabla }_{p_2}\right.
$$
$$
-i^2e_1^2{\cal{E}} _i{\cal{E}} _j\frac{\left( 2E_1+E_2-p_0\right)
\left( 2m_1^2\delta_{ij}-3p_{1i}p_{1j}\right) }{8E_1^6}
$$
$$
\left. -i^2e_2^2{\cal{E}} _i{\cal{E}} _j\frac{\left(
E_1+2E_2-p_0\right) \left( 2m_2^2\delta_{ij}-3p_{2i}p_{2j}\right)
}{8E_2^6}\right\} {\Psi } \left( \stackrel{\rightarrow
}{p_1}\stackrel{\rightarrow }{p_2}\right);
$$
\begin{equation}
V {\Psi } \left( \stackrel{\rightarrow
}{p_1}\stackrel{\rightarrow }{p_2}\right)=\int V\left(
\stackrel{\rightarrow }{p_1}\stackrel{\rightarrow }{p_2};
\stackrel{\rightarrow }{p_1}^{\prime }\stackrel{\rightarrow
}{p_2}^{\prime }\right) {\Psi } \left( \stackrel{\rightarrow
}{p_1}^{\prime }\stackrel{\rightarrow }{p_2} ^{\prime }\right)
d\stackrel{\rightarrow }{p_1}^{\prime }d\stackrel{ \rightarrow
}{p_2}^{\prime },
\end{equation}
where $V $ is the effective potential of the quark interaction.

As follows from the relativistic equation (\ref{usa15}) and
perturbation theory, the correction to the energy of the ground state
is proportional to the second order of the
electrostatic field strength is defined by:
\begin{equation}
\triangle E^{(2)}_n = -2 \pi \alpha \stackrel{\rightarrow
}{{\cal{E}}}^2 = \left\langle n\right| \hat {W}^{\left( 1\right)
}\left| \varphi _n^{\left( 1\right) }\right\rangle +\left\langle
n\right| \hat {W}^{\left( 2\right) }\left| n\right\rangle.
\label{usa16}
\end{equation}
In expression (\ref{usa16}) $\alpha$ is the electrical
polarizability; $\hat{W}^{\left( 1\right) }$ and $\hat{W}^{\left(
2\right) }$ are the matrix elements of operators of the perturbation,
corresponding to the first and second approximations on
the field strength, which are defined in equation (\ref{usa15}); $
\varphi_n ^ {\left (1\right)}$ is the wave function satisfying
equation (\ref{usa15}) in the first order on the value of
$ {\vec {\cal {E}}} $. The calculation of a
wave function $ \varphi _n ^ {\left (1\right)} $ is possible for
defined potentials of interaction between quarks. As it is
visible from equation (\ref{usa16}), the first term contains
nonrelativistic expression of the polarizability. The second term
describes the
relativistic corrections. Calculations
using Coulomb or oscillator potentials show
that these contributions are essential.

The inverse time ordered  Green's function is defined in a
covariant three-dimensional formalism using the diagram approach.
The equation of motion of a two-quark composite system in an
external electromagnetic field up to the second order in the field
strength is obtained. The polarizability of the $ \pi$-meson is
calculated using this formalism. The mean square radius of the
two-quark system is calculated from the wave functions of the
system. In addition, radiative corrections are calculated for both
the  polarizability and the mean square radius.


\begin{thebibliography}{99}

\bibitem{usa1} Weise W. QCD aspects of hadron physics // arXiv: hep-ph/9904487.v1,
1999.
\bibitem{usa2} Maksimenko N.V., Shulga S.G. Relativistic
"trembling" effect of quarks in electric polarizability of mesons
// Journal of nuclear physics. -- 1993. -- Vol. 56. -- ¹ 6. -- P. 201-210 (in
Russian).
\bibitem{usa6}Lucha W., Schoberl F.F. Electric polarizability of mesons
in semirelativistic quark models // arXiv: hep-ph/0204325 v1,
2002.
\bibitem{usa7} Lee R.N., Milstein A.J., Schumacher M.
Electric polarizabilities of proton and neutron and the
relativistic center-of-mass coordinate //  arXiv: hep-ph/0203100,
2002.
\bibitem{usa5}Faustov R.N. Magnetic moment of the relativistic composite system
// Nuovo Cimento. -- 1970. -- Vol. 69A. -- ¹ 1. -- P. 37-46.
\bibitem{usa3} Electromagnetic meson form factors in a
covariant Salpeter model. C.R.Muns, J.Resag, B.C.Metsch,
H.R.Petry // Phys.Rev.C. -- 1995. -- Vol. 52. -- ¹ 4. -- P.
2110-2119.
\bibitem{usa9} Maksimenko N.V., Deryuzhkova O.M., Lukashevich S.A.
The electromagnetic characteristics of hadrons in the covariant
Lagrangian approach // The actual problems of particle physics:
Proceeding of International School-Seminar HEP'01, Gomel, August
7-16, 2001. / JINR. E1, 2-2002-166. -- Dubna, 2002. -- P. 145-156.
\bibitem{usa8} Petrun'kin V.A. Electric and magnetic
polarizabilities of hadrons // Physics of elementary particles
and atomic nuclei. -- 1981. -- Vol. 12. -- ¹ 3. -- P. 692-753 (in
Russian).
\bibitem{usa4} Landau L.D., Lifshitz E.M. Quantum mechanics. --
1989. -- Moscow Science.
\end{thebibliography}
\end{document}